\documentclass[11pt]{article}
\usepackage{moriond,epsfig}

\def\be{\begin{equation}}
\def\ee{\end{equation}}
\def\bea{\begin{eqnarray}}
\def\eea{\end{eqnarray}}

\begin{document}
\makeatletter
\def\fmslash{\@ifnextchar[{\fmsl@sh}{\fmsl@sh[0mu]}}
\def\fmsl@sh[#1]#2{%
  \mathchoice
    {\@fmsl@sh\displaystyle{#1}{#2}}%
    {\@fmsl@sh\textstyle{#1}{#2}}%
    {\@fmsl@sh\scriptstyle{#1}{#2}}%
    {\@fmsl@sh\scriptscriptstyle{#1}{#2}}}
\def\@fmsl@sh#1#2#3{\m@th\ooalign{$\hfil#1\mkern#2/\hfil$\crcr$#1#3$}}
\makeatother
\vspace*{4cm}
\title{SYMMETRIES, MIRCOCAUSALITY AND PHYSICS ON CANONICAL NONCOMMUTATIVE SPACETIME}

\author{ XAVIER CALMET}

\address{
Service de Physique Th\'eorique, CP225 \\
Boulevard du Triomphe \\
B-1050 Brussels \\
Belgium \\
email: xcalmet@ulb.ac.be
}

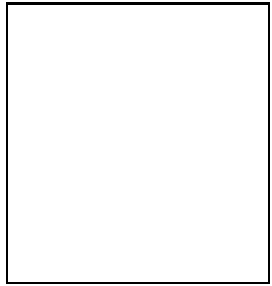
\maketitle\abstracts{
In this paper we describe how to implement symmetries on a canonical noncommutative spacetime. We focus on noncommutative Lorentz transformations. We then discuss the structure of the light cone on a canonical noncommutative spacetime and show that field theories formulated on these spaces do not violate mircocausality.
}

A natural way to implement the notion of minimal length \cite{minlength1} in gauge theories and gravitational theories is to formulate these models on a noncommutative spacetime. The aim of this paper is to reconsider the bounds on spacetime noncommutativity. We shall be dealing with the simplest example one can think of, the canonical noncommutative spacetime. The history of spacetime noncommutativity is not new and was first discussed by Snyder \cite{Snyder:1947qz}  in the early days of quantum field theory at a time where these theories were still plagued by infinities. The motivation to consider spacetime noncommutativity was that introducing a cutoff could help to deal with infinities. Nowadays we know that the quantum field theories relevant for the electroweak and strong interactions are renormalizable and thus cutoff independent, but modifying spacetime at short distance might help for quantum gravity, whatever this theory might be.
 
It is also well-known that non-commuting coordinates are relevant to nature, as soon as one restricts a system to the first Landau level. A textbook example is an electron in a strong magnetic field. Another motivation comes from the fact that non-commuting coordinates naturally appear in string theory \cite{Douglas:2001ba}, although in that case the situation is quite different, since in string theory, the fundamental spacetime is really commutative, only the low energy effective field theory on the brane is described in terms of non-commuting coordinates.

From our point of view spacetime noncommutativity is an extension of quantum mechanics. We  extend the Heisenberg algebra with the noncommutative algebra
\begin{eqnarray} \label{NCA}
[ \hat x^\mu,\hat x^\nu]=i\theta^{\mu\nu}
\end{eqnarray}
where $\theta^{\mu\nu}$ is an antisymmetric constant tensor and $\mu,\nu$ run from 0 to 3. This tensor has mass dimension minus two.

The main difficulty in formulating gauges theories and gravity  on a canonical noncommutative spacetime is to understand how to implement symmetries. In order to have SU(N) gauge symmetries and arbitrary representations for the fields, one needs to consider gauge transformations and fields in the enveloping algebra \cite{Madore:2000en,Jurco:2000fs,Jurco:2000ja,Jurco:2001rq,Calmet:2001na}. One can then map the noncommutative action to an effective action formulated on a commutative spacetime. The dimension four operators are those of a regular Yang-Mills theory and the special nature of spacetime is encoded in the higher dimensions operators. This effective action can be calculated to any order in the tensor $\theta^{\mu\nu}$.  Recently the second order expansion has been performed \cite{Moller:2004qq}. It has been pointed out that these actions are invariant under  noncommutative Lorentz transformations \cite{Calmet:2004ii} which also preserve the algebra (\ref{NCA}). Similar techniques have been applied to formulate Einstein's theory of gravitation on a noncommutative spacetime \cite{Calmet:2005qm}. This leads to a relation between the cosmological constant and the scale of spacetime noncommutativity \cite{Calmet:2005mc}.

There are different type of bounds on the noncommutative nature of spacetime which turn out to be fairly weak when the enveloping algebra approach is used. It is important to differentiate between bounds on new interactions allowed if spacetime is noncommutative, like e.g. new interactions between the known gauge bosons and bounds on Lorentz invariance.  The algebra  (\ref{NCA}) breaks Lorentz symmetry explicitly since $\theta^{\mu\nu}$ is constant and Lorentz invariance is clearly not a symmetry of the action. On the other hand as shown in ref.\cite{Calmet:2004ii}, noncommutative actions are invariant under noncommutative Lorentz transformations. When performing a theoretical calculation e.g. in the case of an atomic clock, one should use a noncommutative Lorentz transformation to go from one coordinate system to another. This implies that the bounds obtained using a regular Lorentz transformation will be affected.  It is thus best not to have to rely on a coordinate transformation to test the noncommutative nature of spacetime but rather to search directly for new vertices that would be a clear signal of spacetime noncommutativity. The bounds obtained by the OPAL collaboration \cite{Abbiendi:2003wv} for the scale of spacetime noncommutativity coming from searches for three photons interactions are only of 141 GeV. Bounds coming from test of Lorentz violation, which once again should be taken with great care are in the 10 TeV range. 

The  relevant bounds  for the enveloping algebra approach are in the TeV region \cite{Calmet:2004dn,Carroll:2001ws,Carlson:2001sw}. The conclusions of ref. \cite{Calmet:2004dn} where it was shown that the relevant bounds are of the order of a TeV have now been strengthen by the proof that the actions obtained in the enveloping algebra approach are renormalizable at least at  one loop \cite{Calmet:2006zy}.  It was shown that physical observables can be calculated at the quantum level
independently of a cutoff. For example the anomalous
magnetic moment of a fermion has the usual quantum electrodynamics
value (more precisely the noncommutative contribution is spin independent and will thus not contribute to the anomalous magnetic moment  when measured with classical methods). 
The value of the $\beta$-function is nevertheless an issue:
\begin{equation}
 \beta(g) = \frac{1}{g} Q \frac{dg}{dQ}
  = -\left(
       \frac{22}{3} - \frac{4}{3} 
     \right)\, \frac{g^2}{16\pi^2},
   \label{beta}
\end{equation}
 where the contribution $22/3$ is due to the structure of the gauge interaction which is similar
to that of the nonabelian SU(2) Yang-Mills theory  and the factor $4/3$ is due to the fermion field $\hat \psi$ which is in the enveloping algebra.  We see that the value of the $\beta$-function  does not match that of quantum electrodynamics on a commutative spacetime.  It is however possible to
assume that the noncommutative parameter $\theta^{\mu\nu}$ is not
a simple constant but is spacetime or energy dependent as studied in ref. \cite{Calmet:2003jv}. The assumption that $\theta^{\mu\nu}$ is scale dependent is not that far-fetched, indeed if it is the expectation value of a background field, as e.g. in the string theory picture \cite{Seiberg:1999vs}, one would expect a scale dependence of the renormalized expectation value. 

 It thus remains important to search for modifications of the gauge bosons sector due spacetime noncommutativity at the LHC or future linear colliders  \cite{Behr:2002wx,Duplancic:2003hg,He:2006yy} or in other low energy experiments  \cite{Melic:2005su,Melic:2005hb} as these bounds are independent of a cutoff and thus very clean.

One of the main motivations to consider gauge theories or gravity on a noncommutative spacetime is that it introduces a fundamental length in these theories. If this length is a physical observable it should not depend on the observer and thus coordinate transformations should leave invariant the noncommutative algebra.  We now study further the noncommutative Lorentz symmetry introduced in ref. \cite{Calmet:2004ii}.

Let us consider the noncommutative algebra once again:
\begin{eqnarray} \label{NCA2}
[ \hat x^\mu,\hat x^\nu]=i\theta^{\mu\nu}
\end{eqnarray}
where $\mu,\nu$ run from 0 to 3. 
Furthermore one has the Heisenberg algebra:
\begin{eqnarray}
[ \hat x^\mu,p^\nu]=i \hbar \eta^{\mu\nu}
\end{eqnarray}
(we have introduced a noncommutative relation for time and energy \footnote{another option is to set $\theta^{0i}=0$ as done in ref. \cite{Calmet:2004ii}.}) and
\begin{eqnarray}
[  p^\mu,p^\nu]=0.
\end{eqnarray}

It is easy to see that one can introduce a new operator $x_c^\mu$ defined by
\begin{eqnarray} \label{CLT}
x_c^\mu=\hat x^\mu +\frac{1}{2\hbar} \theta^{\mu\nu} p_\nu,
\end{eqnarray}
 which leads to the following algebras:
\begin{eqnarray} \label{NAlg}
[ x_c^\mu, x_c^\nu]=0, \ [  x_c^\mu,p^\nu]=i \hbar \eta^{\mu\nu} \ \mbox{and} \ [  p^\mu,p^\nu]=0,
\end{eqnarray}
i.e. $x_c^\mu$ are commuting coordinates.   Given the algebras (\ref{NAlg}), we can define transformations
\begin{eqnarray} 
x_c^{\mu\prime}= \Lambda^\mu_{\ \nu} x_c^\nu
\end{eqnarray}
that leave the interval $s^2= \eta_{\mu\nu} x_c^\mu x_c^\nu$
invariant if $\eta_{\mu\nu} \Lambda^\mu_{\ \alpha} \Lambda^\nu_{\
\beta} = \eta_{\alpha \beta} $. Note that $p^\mu$ transforms as an
usual Lorentz vector, i.e.
\begin{eqnarray} 
p^{\mu\prime}= \Lambda^\mu_{\ \nu} p^\nu.
\end{eqnarray}
One thus finds that the transformation (\ref{CLT}), implies that $\hat x^\mu$ transforms as
\begin{eqnarray} \label{NLT}
\hat x^{\mu\prime} &=& \Lambda^\mu_{\ \nu} \hat x^\nu + \frac{1}{2\hbar}   \Lambda^\mu_{\ \nu} \theta^{\nu \rho} p_\rho 
- \frac{1}{2\hbar}  \theta^{\mu\nu} \Lambda_\nu^{\ \rho} p_{\rho} 
\end{eqnarray}
which defines the noncommutative Lorentz transformation. 
The square of the invariant length for the commutative coordinate $x^\mu_c$ is
\begin{eqnarray} 
s^2= \eta_{\mu\nu} x^\mu_c x^\nu_c.
\end{eqnarray}
Using the variable transformation (\ref{CLT}), one finds that the square of the noncommutative invariant length is given by
\begin{eqnarray}  \label{invlen}
s^2_{nc}= \hat x^\mu \hat x_\mu + \frac{1}{\hbar} \theta_{\mu\nu} \hat
x^\mu p^\nu + \frac{1}{4\hbar^2} \theta^{\mu\alpha} \theta_{\mu \beta}
p_\alpha p^\beta.
\end{eqnarray}
It is easy to verify that $s^2_{nc}$ is left invariant by the noncommutative Lorentz transformation 
(\ref{NLT}). This is the way we define the noncommutative Lorentz transformations, those are the transformations that leave $s^2_{nc}$ invariant. Note that the Lie algebra  of the Lorentz group is not deformed \footnote{it should be stressed that this symmetry is not a deformed symmetry. Deformed symmetries correspond to explicit symmetry breaking \cite{Gonera:2005hg} and are thus not  physical symmetries.}. It is easy to verify that this symmetry is a symmetry of the actions discussed previously in the literature whether using the Lie algebra approach (see e.g. ref. \cite{Douglas:2001ba}) or the enveloping algebra approach \cite{Jurco:2000ja}.

Derivatives have to be defined in such a way, that they do
not lead to new relations for the coordinates. In the canonical case,
it is easy to show that $\hat x^\alpha - i
\theta^{\alpha\rho}\hat\partial_\rho$ with $\hat \partial_\rho \hat
x^\mu=\delta^\mu_\rho +\hat x^\mu \hat \partial_\rho$ commutes with
all coordinates \cite{Jurco:2000ja}. One thus finds $\theta^{\nu \mu} \hat \partial_\mu f=
-i  [\hat x^\nu,f]$. In our case we need a derivative
which is compatible with the noncommutative Lorentz symmetry. We
define the derivative in the following way:
\begin{eqnarray}
i \theta_{\mu\nu} \hat \partial^\nu f = 2 [ \hat x_\mu +\frac{1}{2 \hbar} \theta_{\mu\alpha} p^\alpha, f ]
\end{eqnarray}
with $[p^\mu,f]=-i\hbar \partial^\mu f$. Note that the left hand side
of the equation is covariant. One finds that the derivative
$\hat \partial_\nu$ transforms as
\begin{eqnarray}
 \theta^{\alpha \nu} \hat \partial_\nu^\prime= \Lambda^{\alpha}_{\ \beta} \theta^{\beta \rho} \hat \partial_\rho
\end{eqnarray}
under a noncommutative Lorentz transformation.  

The noncommutative Lorentz transformation is compatible with gauge
transformations. Remember that one has to introduce a covariant
coordinate $\hat X^\mu$ \cite{Madore:2000en} such that $\hat \delta_{\hat
\Lambda} (\hat X^\mu \hat \Psi(\hat x))=\hat \Lambda \hat X^\mu \hat \Psi(\hat
x)$ where $\hat \Lambda$ is a noncommutative gauge transformation. One
finds that $\hat X^\mu=\hat x^\mu+\hat B^\mu$ with $\hat \delta_{\hat
\Lambda} \hat B^\mu=i [\hat \Lambda, \hat B^\mu] -i[\hat x^\mu, \hat
\Lambda]$.  The Yang-Mills gauge potential $\hat A^\mu$ is related to
the gauge potential for the coordinate $\hat B^\mu$ by the relation
$\hat B^\mu=\theta^{\mu \nu} \hat A_\nu$ and the covariant derivative
$\hat D^\mu$ is given by $\theta^{\nu\mu}  \hat D_\mu=-i \hat
X^\nu $. The coordinate gauge potential $\hat B_\mu$ transforms
as $\hat B^\prime_\mu(\hat x^\prime) = \Lambda_\mu^{\ \nu} \hat B_\nu(\hat x)$, one thus finds
that the noncommutative Yang-Mills potential transforms as 
 \begin{eqnarray}
\theta^{\nu\mu} \hat A^\prime_\mu(\hat x^\prime) =  \Lambda^\nu_{\ \rho} \theta^{\rho
\sigma} \hat A_\sigma(\hat x). 
\end{eqnarray}
The noncommutative covariant derivative transforms as
\begin{eqnarray} \label{T1}
\theta^{\rho\mu}  \hat D_{\mu}^\prime= \Lambda^\rho_{\ \sigma}
\theta^{\sigma \alpha} \hat D_{\alpha}
\end{eqnarray}
under a noncommutative Lorentz transformation.
 The field strength $\hat F_{\mu\nu}$ is given by  $\hat F_{\mu\nu}= i [\hat D_\mu,\hat D_\nu]$, it  transforms as 
\begin{eqnarray} \label{T2}
 \theta^{\rho\mu}   \theta^{\kappa\nu}  \hat F_{\mu\nu}^\prime(\hat x^\prime)=
\Lambda^\rho_{\ \sigma} \theta^{\sigma \alpha}
                             \Lambda^\kappa_{\ \xi} \theta^{\xi \beta}  \hat F_{\alpha\beta}(\hat x)
\end{eqnarray}
under a noncommutative Lorentz transformation.
The noncommutative spinor field $\hat \Psi$ transforms as 
\begin{eqnarray} \label{T3}
\hat \Psi^\prime(\hat x^\prime)= \exp \left ( -\frac{i}{2} w^{\alpha \beta} S_{\alpha \beta} \right  )  \hat \Psi(\hat x),
\end{eqnarray}
with $S^{\mu\nu}=\frac{i}{4}[\gamma^\mu,\gamma^\nu]$.  Note that if
the fields are taken in the enveloping algebra, the leading
order field of the Seiberg-Witten expansion,
i.e. the classical field, also transforms according to (\ref{T1}),
(\ref{T2}) and (\ref{T3}).

Given (\ref{T1}),
(\ref{T2}) and (\ref{T3}) is it easy to verify that the effective
action obtained in the leading order in $\theta$ after the expansion
of the noncommutative fields via the Seiberg-Witten map and of the
star product:
\begin{eqnarray} \label{action}
S &=& \int d^4 x \bigg[ \bar \psi (i \! \not \! \! {D} - m
)\psi - \frac 1 4 \, \theta^{\mu \nu} \bar \psi F_{\mu \nu}
(i \not \! \! {D} - m )\psi - \frac{1}{2} \, \theta^{\mu \nu}
\bar \psi \gamma^\rho F_{\rho \mu} \, i {D}_\nu \psi
\\ \nonumber &&
 -\frac{1}{2} \, {\rm Tr} \, F_{\mu \nu} F^{\mu \nu}
+ \frac{1}{4}  \, \theta^{\mu \nu} \, {\rm Tr} \, F_{\mu \nu}
F_{\rho \sigma} F^{ \rho \sigma} -  \, \theta^{\mu
\nu} \, {\rm Tr} \, F_{\mu \rho} F_{\nu \sigma} F^{ \rho
\sigma} \bigg] + {\cal O}(\theta^2)
\end{eqnarray}
is invariant under noncommutative Lorentz transformations.  

We now have all the required tools to study the light cone of a photon described by QED formulated on a noncommutative spacetime which is invariant under the noncommutative Lorentz invariance. The light cone is defined by setting the invariant length defined in (\ref{invlen}) equal to zero, i.e. $s^2_{nc}=0$. We find
\begin{eqnarray}  \label{lightcone}
0= (\hat x^\mu +\frac{1}{2\hbar} \theta^{\mu\rho} p_\rho)(\hat x_\mu +\frac{1}{2\hbar} \theta_{\mu\sigma} p^\sigma)=\hat x^\mu \hat x_\mu + \frac{1}{\hbar} \theta_{\mu\nu} \hat
x^\mu p^\nu + \frac{1}{4\hbar^2} \theta^{\mu\alpha} \theta_{\mu \beta}
p_\alpha p^\beta,
\end{eqnarray}
which is not $0=\hat x^\mu \hat x_\mu$, which would have been expected if one had tried to implement Lorentz invariant in the usual way. We can now reconsider the result  obtained in ref. \cite{greenberg} where the commutator  between $:\phi(x) \star \phi(x):$ and $\partial^y_\mu : \phi(y) \star \phi(y):$ has been calculated. The author of ref. \cite{greenberg}   finds that this commutator is given by
\begin{eqnarray} 
\sum_{s=\pm1, t=\pm1}\delta(x^1-y^1-s \theta(p_2+t p_2')) \delta(x^2-y^2-s \theta (p_1+t p_1')) \delta(x^3-y^3),
\end{eqnarray}
and concludes that microcausality is violated, which would be the case if the light-cone was defined as $\hat x^\mu \hat x_\mu=0$. However, we see that this result corresponds precisely to the light cone defined in  (\ref{lightcone}) and we thus argue that the theory is not violating microcausality. We want to had a remark concerning the unitarity issue of noncommutative gauge theories. It is well known that in the Lie algebra approach, unitarity is violated if $\theta^{0i}\neq 0$ \cite{Gomis:2000zz}, but in the case of the enveloping algebra approach (approach we used in deriving the standard model on a noncommutative spacetime \cite{Calmet:2001na}) we have to deal with an effective field theory and it is not trivial to discuss unitarity in that framework. Indeed, let us imagine that unitarity is violated at an energy scale $\Lambda_{nc}$, but this is the scale where higher order operators suppressed by powers of $\Lambda_{nc}$ become relevant. It is thus not clear that there is a problem with unitarity within the enveloping algebra approach and we thus do not set  $\theta^{0i}= 0$. 

In this paper we have shown that gauge theories formulated on noncommutative spaces obey microcausality when the light-cone is defined in accordance  with the noncommutative Lorentz invariance which is the exact spacetime symmetry of these models. Furthermore the bounds on the noncommutative scale are only of the order of a TeV. The enveloping algebra approach is thus in a very good shape. 

\section*{Acknowledgments}
This work was supported in part by the IISN and the Belgian science
policy office (IAP V/27).

\end{document}